\documentclass{czjphys}         
\usepackage{amsfonts}
\begin{document}
\title{On the uniqueness of harmonic coordinates\footnote{Dedicated to Prof. Ji\v{r}\'{\i} Hor\'{a}\v{c}ek on the occasion of his 60th birthday. Undoubtedly, Ji\v{r}\'{\i} has been using harmonic coordinates---in flat space these are the Cartesian coordinates---in his quantum calculations happily for years. We hope that his sense for harmony in music and life will also appreciate the harmonic gauge in curved spacetimes...}}  
%
\authori{Ji\v{r}\'{\i} Bi\v{c}\'ak}      \addressi{Institute of Astronomy, The Observatories, Cambridge CB3 0HA, UK\\
and\\
Institute of Theoretical Physics,  Faculty of Mathematics and Physics, Charles University,\\
  V Hole\v{s}ovi\v{c}k\'{a}ch 2, 180 00 Prague 8, Czech Republic}
\authorii{Joseph Katz}     \addressii{Institute of Astronomy, The Observatories, Cambridge CB3 0HA, UK\\
and\\
Racah Institute of Physics, The Hebrew University, Jerusalem 91904, Israel}
\authoriii{}    \addressiii{}
\authoriv{}     \addressiv{}
\authorv{}      \addressv{}
\authorvi{}     \addressvi{}
%
\headauthor{J. Bi\v{c}\'ak and J. Katz}            
\headtitle{On the uniqueness of harmonic coordinates}             
\lastevenhead{J. Bi\v{c}\'ak and J. Katz: On the uniqueness of harmonic coordinates} 
\refnum{A}
\daterec{XXX}    
\issuenumber{5}  \year{2005}
\setcounter{page}{105}
\maketitle

\begin{abstract}
Harmonic coordinate conditions in stationary asymptotically flat spacetimes with matter sources have more than one solution. The solutions depend on the degree of smoothness of the metric and its first derivatives, which we wish to impose across the material boundary, and on the conditions at infinity and at a suitable point inside the matter. This is illustrated in detail by simple fully solvable examples of static spherically symmetric spacetimes in global harmonic coordinates. Examples of stationary electrovacuum spacetimes described simply in harmonic coordinates are also given. They can represent the exterior fields of material discs.

The use of an appropriate background metric considerably simplifies the calculations.
\end{abstract}

\section{Introduction}
Harmonic coordinates have been frequently used in general relativity even before the publication of the monograph \cite{FO} by their main protagonist. They present the straightforward generalisation of the Lorentz gauge in electrodynamics and play a substantial role in approximation methods dealing with motion and radiation; they allow one to develop a direct post-Newtonian iteration of the equations of motion (see \cite{BL} and \cite{BD} for reviews including the most recent applications). The harmonic coordinates have been also frequently employed in numerical relativity (see, e.g., \cite{W}, \cite{LE}). The Cauchy problem for the Einstein equations was first rigorously tackled in harmonic coordinates (see \cite{FR} for a comprehensive review) and, quite surprisingly, the global stability (for restricted data) of Minkowski space for the Einstein vacuum equations was recently proven in harmonic coordinates \cite{LR}, providing so a considerably simpler approach to the proof of the celebrated result by Christodoulou and Klainerman.

In view of numerous roles the harmonic coordinates play in vacuum regions it is  surprising how very few examples of spacetimes with material sources are described globally in these coordinates. In 1998, Q.-H. Liu \cite{Li1} found a harmonic coordinate system for the Schwarzschild interior solution, i. e., for a star of incompressible fluid (see, e. g., \cite{We}, p. 330). Although he limited the problem to a special (not very physical) situation in which the central pressure becomes infinite, the harmonic radial coordinate in the interior is given in terms of a hypergeometric function of the Schwarzschild radial coordinate, hence, as an infinite series. When transforming the exterior from the Schwarzschild coordinates $\{ t, r, \theta, \varphi \}$  to the harmonic coordinates $\{ X^{\mu} \}$, where
\begin{equation}\label{1}
X ^0 = t, \;\;\;\; X ^1 = R(r) \sin \theta \cos \varphi , \;\;\; X^2 = R(r) \sin \theta \sin \varphi , \;\;\; 
X^3 = R(r) \cos \theta ,
\end{equation}
Liu wrote down the new radial (``harmonic'') coordinate $R$ in full generality, determined in terms of two arbitrary constants $C$ and $K$ ($M$ is the Schwarzschild mass):
\begin{equation}\label{2}
R(r) = C(r-M) + K \left[ \frac{r-M}{2M} \ln(1-\frac{2M}{r}) + 1 \right].
\end{equation}
Since one is usually interested in the vacuum region, it is customary to consider one convenient solution only (cf. \cite{We}, p. 181), putting $K = 0$ and $C = 1$; then the harmonic coordinates become Minkowskian at (asymptotically flat) infinity. Liu also puts $C = 1$, but leaves $K$ to be determined by the junction conditions at the surface of the star, namely, by the requirement that the derivative $dR/dr$  be continuous across the surface. (In Liu's specific case, $K \approx 0.06 r_0$, where $r_0$ is the Schwarzschild radius of the sphere.) Gergely \cite{Ge} rightly argued that the continuity of the derivative is \textit{not} required by the junction conditions since the matching only requires that the induced metric and the extrinsic curvature at the junction hypersurface  $(r = r_0)$ be the same from both sides, i.e., $g_{RR}$ need not be continuous in the spherical case. Hence, one can put $K = 0$ in Eq. (\ref{2}); Liu admits this in his reply \cite{Li2} to Ref. \cite{Ge}.

However, as Liu \cite{Li2} and, in particular, the authors of Refs. \cite{AMR}, \cite{HMR}, \cite{MMR} recently accentuated, one may take the view that the matching is ``completely solved'' only when ``admissible coordinates in the sense of Lichnerowicz \cite{LW}'' are found: in such coordinates the first derivatives of all metric coefficients should also be continuous across the matching  surface. If one wishes to use harmonic coordinates for this purpose, a constant $K \neq 0$ in (\ref{2}) is in general needed. In fact, in \cite{HMR} the authors generalised Liu's result to finite central pressures. To find harmonic coordinates in such a general case of a star with a uniform density one has to  solve the Heun differential equation of second order with 4 regular singular points \cite{HEU}. The solutions are then given in terms of infinite series which are still more complicated than in Liu's work.

Turning to more general cases, we are not aware of any work in which a \textit{stationary} exact spacetime with a source is described explicitly in harmonic coordinates in a tractable form. In \cite{AMR}, for example, only the asymptotic expansion of the stationary Kerr metric in harmonic coordinates is derived to order five in $R^{-1}$, and it is compared with a similar expansion of the static (but not spherically symmetric) Curzon metric.

In the present note we first show how harmonic gauge conditions can be written employing an auxiliary background metric which allows to take full advantage of the symmetries of spacetime instead of going first to Minkowski-like coordinates $\{ X^\mu \} = \{ T, X, Y, Z \}$. We then give simple examples which represent material sources and their fields both being expressed in globally smooth harmonic coordinates. As we shall see in detail, there is more than one possible choice. Different smooth harmonic coordinates relate to how much smoothness we ask of the metric components across the boundary of the matter and at the centre (in case of spherical symmetry). Fock \cite{FO} in advocating harmonic coordinates and their uniqueness was not concerned with such considerations.

The examples include the ``conformastatic metrics'' describing static charged dust and its electric and gravitational fields with the electric and gravitational forces in exact balance (see, e.g., \cite{SKM}, \cite{IVA} and references there in). Using a specific example of an interior solution \cite{Bo1}, we show that, in addition to the ``natural'' set of global harmonic coordinates, there exists another ``unconventional'' set. These alternative harmonic coordinates are ``admissible in the sense of Lichnerowicz'', however, their origin is ``shifted''  to physical infinity and infinity becomes the origin. In the static case, we also generalise Liu's result (\ref{2}) to the electrovacuum Reissner-Nordstr\"{o}m metrics with arbitrary $M$ and $e$. 

Next, we consider stationary spacetimes. We show that axially symmetrical \textit{conformastationary} electrovacuum metrics, for which disc sources can be found \cite{KBL1}, can be described explicitly in harmonic coordinates in a surprisingly simple form. As a special case, the conformastationary solutions contain the Kerr-Newman metrics with $e^2 = M^2$ but a general $a$ ($e$ being the total charge and $aM$ the angular momentum). Hence, there exists a family of Kerr-Newman metrics, which do not represent black holes but can be produced by charged, rotating discs, that can be simply described in harmonic coordinates. In fact, Kerr metric itself (even for $a^2<M^2$) can also be transformed into harmonic coordinates \cite{Ru}, \cite{AN} but it acquires a complicated form that is hardly useful in analytic problems. Nevertheless, the disc sources can be constructed for the Kerr metric as well \cite{BiL}, \cite{PiL}. 

\section{The generalised harmonic gauge conditions}
In our recent work on suitable mappings of the physical spacetime with metric ${\bf g}$ onto an auxiliary background spacetime with metric ${\bf \bar{g}}$ which would indicate how the total KBL energy \cite{KBL2} might be split into the matter and gravitational parts, the harmonic mappings appear to play a role. We needed to describe not only vacuum regions but also those containing matter in harmonic coordinates. This issue, in fact, stimulated the present remarks. In the formulations of conservation laws in general relativity the choice of an appropriate background (at least in asymptotic regions) is important. A suitable background metric, not necessarily in Minkowski coordinates even with the background being flat, can considerably simplify the harmonic gauge conditions.

In standard textbooks (see, e.g., \cite{FO}, \cite{We}), harmonic coordinates $X^\mu$ are defined by the conditions
\begin{equation}\label{3}
\frac{1}{\sqrt{-g}}\frac{\partial}{\partial X^\nu}(\sqrt{-g} g^{\mu\nu}) = 0.
\end{equation}
As a consequence of Eq. (\ref{3}), each of the harmonic coordinates $X^\mu$ satisfies the wave equation $\square X^\mu = 0$ where the invariant d'Alembertian is given by
\begin{equation}\label{4}
\square \psi \equiv \sqrt{-g} \frac{\partial}{\partial X^\nu} \left( \sqrt{-g} g^{\nu\mu}\frac{\partial \psi}{\partial X^\mu} \right)
\end{equation}
for any scalar $\psi$. In the mathematical literature (see, e.g., \cite{LR}) $X^\mu$  are thus called ``wave coordinates''. One of the well-known advantages of harmonic coordinates is a simplified form of the covariant wave equation in a curved spacetime---it only contains the second derivatives:
\begin{equation}\label{5}
g^{\mu\nu} \frac{\partial^2\psi}{\partial X^\mu\partial X^\nu} = 0,
\end{equation}
as follows from (\ref{3}) and (\ref{4}).

Harmonic coordinates are most often used in asymptotically flat spacetimes representing insular systems, hence, it is commonly assumed that they go over to Minkowskian coordinates at infinity. However, in case of physical systems with some symmetry the spacetime metrics are naturally given in terms of the coordinates reflecting the symmetry. In order to see how such metrics can be converted to harmonic coordinates, it is advantageous to introduce a suitable background spacetime $\overline{\mathcal{M}}$ with a background metric ${\bf \overline{g}}$, and rewrite harmonic conditions (\ref{3}) in a form invariant with respect to the coordinate transformations in $\overline{\mathcal{M}}$. (Clearly, conditions (\ref{3}) are not form-invariant with respect to coordinate changes in the \textit{physical} spacetime $\mathcal{M}$ because they restrict the coordinates in $\mathcal{M}$.) For the purpose of the present note in which asymptotically flat spacetimes $\mathcal{M}$ are considered, it is natural to choose $\overline{\mathcal{M}}$ flat; in a cosmological context, a suitable Friedmann-Robertson-Walker universe may be an appropriate choice (cf., e.g., \cite{KBL2}). We shall assume that there is a region $\mathcal{D}$ in $\mathcal{M}$, covering a neighbourhood of infinity, the points $P$ of which can be mapped by the 1-1 mapping onto the points $\overline{P}$ in $\overline{\mathcal{M}}$. Assume that in $\mathcal{M}$ harmonic coordinates $\{X^\mu\}$ satisfying (\ref{3}) are introduced. They are going over to Minkowski coordinates at infinity. Assume that the mapping between $\mathcal{M}$ and $\overline{\mathcal{M}}$ is such that to a point $P\in \mathcal{D}$ with coordinates $\{X^{\mu}\}$ corresponds $\overline{P}$ in $\overline{\mathcal{M}}$ with Minkowski coordinates having the same values, $\overline{X}^\mu_{\overline{P}} = X^\mu_P$. Any object like $g_{\mu\nu}(X)$ or $g^{\mu\nu}(X)$ in $\mathcal{M}$ can be brought over to $\overline{\mathcal{M}}$---with the same components. Any coordinate transformation $X^\mu (x^\rho)$ in $\overline{\mathcal{M}}$ induces a transformation in $\mathcal{M}$ which has the same functional form. Such transformations will change the Minkowski metric $\overline{\eta}_{\mu\nu}$ in $\overline{\mathcal{M}}$ into a $\overline{g}_{\mu\nu}(x^\rho)$ with connection coefficients $\overline{\Gamma}^\lambda_{\mu\nu}(x^\rho)$ etc. The difference between Christoffel symbols in $\mathcal{M}$ and $\overline{\mathcal{M}}$, 
\begin{equation}\label{6}
\Delta^\lambda_{\mu\nu} = \Gamma ^\lambda_{\mu\nu} - \overline{\Gamma}^\lambda_{\mu\nu},
\end{equation}
is a tensor, $\sqrt{-g}$ and $\sqrt{-\overline{g}}$ are scalar densities of weight $1$, $\sqrt{-g}/\sqrt{-\overline{g}}$ is a scalar $(g = det |g_{\mu\nu}|$, $\overline{g} = det | \overline{g}_{\mu\nu}| $), etc.---see \cite{KBL2} for more details.

The harmonic conditions (\ref{3}) can easily be converted into a covariant form using the covariant derivative $\overline{\mathcal{D}}_\nu$ with respect to $\overline{g}_{\mu\nu}$:
\begin{equation}\label{7}
\frac{1}{\sqrt{-g}} \overline{D}_\nu (\sqrt{-g} g^{\mu\nu}) = 0.
\end{equation}
Regarding the tensor density character of $\sqrt{-g} g^{\mu\nu}$, and using the explicit expressions for affine connections in terms of metrics and their derivatives, the harmonic conditions (\ref{7}) can be brought into the form
\begin{equation}\label{8}
\frac{1}{\sqrt{-g}} {\overline{D}}_\nu (\sqrt{-g}g^{\mu\nu}) = - g^{\rho\sigma}\Delta^\mu_{\rho\sigma} = 0,
\end{equation}
where $\Delta^\mu_{\rho\sigma}$ is given by (\ref{6}). Equation (\ref{8}) is convenient for direct calculations. This is especially the case when dealing with spherically symmetric spacetimes.

Imagine $\mathcal{M}$ is a general---not necessarily vacuum---spherically symmetric static spacetime with a metric given in the standard Schwarzschild-type coordinates $\{ t, r, \theta, \varphi \} \equiv \{ x^0, x^1, x^2, x^3\}$ as 
\begin{equation}\label{9}
ds^2 = a^2(r) dt^2 - \frac{1}{\alpha^2(r)} dr^2 - r^2 ( d\theta^2 + \sin ^2 \! \theta \; d \varphi^2).
\end{equation}
With $\mathcal{M}$ being spherically symmetric, static, and asymptotically flat, a natural choice of the background metric in $\overline{\mathcal{M}}$ is the flat metric in spherical coordinates,
\begin{equation}\label{10}
d\overline{s}^2 = dt^2 - dR^2 - R^2 (d\theta^2 + \sin ^2 \! \theta \; d \varphi^2).
\end{equation}
Notice that as a consequence of symmetries we at once identified coordinates $t, \theta, \varphi$ in both $\mathcal{M}$ and ${\overline{\mathcal{M}}}$ but not the radial coordinates. Schwarzschild-type $r$ in (\ref{9}) may not, using the standard transformation (\ref{1}), lead to harmonic coordinates $X^\mu$ (it does not), whereas $R$ in (\ref{10}) clearly does since Minkowski coordinates are trivially harmonic. Denoting in $\mathcal{M}$ the radial coordinate by the same $R$ which by (\ref{1}) is associated with harmonic $X^\mu$ in $\mathcal{M}$, we rewrite the metric (\ref{9}) in the form
\begin{equation}\label{11}
ds^2 = a^2 (r(R)) dt^2 - \frac{r'^2}{\alpha^2(r(R))}dR^2 - r^2(R) (d\theta^2 + \sin ^2 \! \theta \; d \varphi^2),
\end{equation}
where prime is $d/dR$. Using (\ref{11}) as $g$ and (\ref{10}) as $\overline{g}$ in the harmonic conditions (\ref{8}), we easily find that they are satisfied for $\mu = 0, 2,3$; for $\mu = 1$ we get a second order differential equation for $R(r)$:
\begin{equation}\label{12}
(r^2 a\alpha) \frac{d^2R}{dr^2} + \frac{d(r^2 a\alpha)}{dr}\frac{dR}{dr} - 2 \frac{a}{\alpha}R = 0.
\end{equation}
This equation coincides with Eq. (8.1.15) in  \cite{We} where it is derived from the standard form (\ref{3}) of the harmonic conditions. It is easy to see that in the case of the vacuum Schwarzschild metric, the expression (\ref{2}) leading to general harmonic coordinates solves Eq. (\ref{12}).

The advantage of using the covariant form (\ref{8}) of the harmonic coordinates will be seen still more clearly in Section 5.

\section{Harmonic coordinates in static, spherically symmetric electrovacuum spacetimes}
As is well known, in the case of static spherically symmetric electrovacuum spacetimes which represent the exterior fields of charged spheres (possibly non-static), charged black holes, or naked singularities, the metric is just the Reissner-Nordstr\"{o}m solution given in Schwarzschild-like coordinates by Eq. (\ref{9}) in which
\begin{equation}\label{13}
a^2 = 1 - \frac{2M}{r} + \frac{e^2}{r^2} = \alpha ^2.
\end{equation}
Equation (12) for $R$, implying the harmonic coordinates by relations (\ref{1}), can be easily solved like in the vacuum Schwarzschild spacetime. However, now three cases arise depending on the relation between the mass $M$ and the charge $e$:

If $e^2 < M^2$, we obtain the general solution in the form
\begin{equation}\label{14}
R = C (r - M) + KM \left[ \frac{r-M}{r_1 - r_2} \ln \left| \frac{1-(r_1/r)}{1-(r_2/r)} \right| +1 \right],
\end{equation}
where $C,K$ are arbitrary dimensionless constants, and $r_{1,2} = M\pm \sqrt{M^2 - e^2}$ are constants determining, in the black-hole case, the locations of outer and inner horizons in Schwarzschild coordinates. (For $e=0$ we get back Eq. (2)). In the special case of $e^2 = M^2$, we find
\begin{equation}\label{15}
R = C(r-M) + K \frac{M^3}{(r-M)^2},
\end{equation}
where $r_1 = r_2 = M$ denotes the location of the (degenerate) horizon of an extreme charged black hole. The horizon does not arise if there is a sphere of charged dust producing the ``extreme'' Reissner-Nordstr\"{o}m metric in the exterior (see next section). For $e^2 > M^2$---corresponding to a naked singularity if no material sphere occurs---the general solution of Eq. (\ref{12}) reads
\begin{equation}\label{16}
R = C(r-M) + KM \left[ \frac{r-M}{\sqrt{e^2-M^2}} \arctan \left( \frac{r-M}{\sqrt{e^2-M^2}}\right) + 1 \right].
\end{equation}

Since at spatial infinity the Schwarzschild-like $r$ implies (by using Eq. (\ref{1})) harmonic coordinates, it is natural to put constant $C=1$ in the first two cases with $e^2\leq M^2$---Eqs. (\ref{14}) and (\ref{15}). However, constant $K$ in (\ref{14}) and (\ref{15}) cannot be fixed by a condition at infinity because it multiplies expressions vanishing at $r\rightarrow \infty$. It can be determined by junction conditions at the surface of a charged sphere, as it is done in \cite{Li1} and \cite{MMR} for the Schwarzschild interior solution. For example, using $K$ one could match the interior solution of Kramer and Neugebauer \cite{KM} to the exterior Reissner-Nordstr\"{o}m metric. Since this solution was generated by an ``invariance transformation'' from the interior Schwarzschild solution, it will, however, be difficult to find the harmonic coordinates in the interior in order to have a global harmonic gauge everywhere. Nevertheless, in the extremal case with $e^2 = M^2$, the situation is enormously simplified.

Before turning to this case, let us note that with $e^2>M^2$ the condition at infinity does influence the value of constant $K$. Requiring $R = r$ at $r\rightarrow \infty$, we can have either $C=1, K = 0$ (the simplest possibility fixing both constants uniquely), or $C + KM (e^2 - M^2)^{-\frac{1}{2}}(\pi/2) = 1$, which gives
\begin{equation}\label{17}
R = r - M + KM \left\{ \frac{r-M}{\sqrt{e^2-M^2}} \left[ \arctan \left( \frac{r-M}{\sqrt{e^2-M^2}}\right) - \frac{\pi}{2} \right] +1 \right\}.
\end{equation}

Again a free constant is available for matching at the charged-sphere boundary.

\section{Spacetimes of electrically counterpoised dust}
Static bodies made of electrically counterpoised dust---in the terminology of Bonnor (see, e.g., \cite{Bo2})---offer simple, physically not unrealistic, models to discuss, for example, large redshifts \cite{Bo2}, the hoop conjecture \cite{Bo3}, or the linear dragging of inertial frames \cite{LBK}. Their global spacetimes can be covered in both exterior and interior regions by harmonic coordinates $X,Y,Z$ in which the metric takes the form
\begin{equation}\label{18}
ds^2 = U^{-2} dt^2 - U^2 (dX^2 + dY^2 + dZ^2),
\end{equation}
where $U = U(X,Y,Z)$. The spacetimes are static but need not have any symmetry. The sources are made of dust whose mass and charge densities $\rho$ and $\sigma$ are equal (with $G=c=1$). The function $U$ satisfies flat-space Laplace's equation in the exterior region $(E)$, $\nabla ^2 U_E = 0$ , whereas in the interior $(I)$ $\nabla ^2 U_I = - 4\pi U_I^3 \rho$. It is not difficult to choose $U_I$ such that $\rho > 0$ and $U_I = U_E$, $U_{I,\mu} = U_{E,\mu}$ on some 2-surface which corresponds to the body's boundary. For example, the spacetime representing a prolate spheroid of electrically counterpoised dust was constructed in this way \cite{Bo3}. The electric field is given by $E = -U^{-2} \nabla U$ (see, e.g., \cite{Bo3}, or Appendix in \cite{LBK}, for more details). The spacetimes described by the metric (\ref{18}) with sources of electrically counterpoised dust offer the simplest examples of globally regular solutions in global harmonic coordinates which are simultaneously ``admissible coordinates in the sense of Lichnerowicz'', as mentioned in the Introduction.

We shall now consider the simplest versions of these spacetimes assuming spherical symmetry. The electrovacuum spacetime outside a sphere of electrically counterpoised dust is described by the extreme Reissner-Nordstr\"{o}m metric given in Schwarz\-schild coordinates by Eq. (\ref{9}) in which $a,\alpha$ are determined by (\ref{13}) with $e^2 = M^2$ so that
\begin{equation}\label{19}
ds^2 = \left( 1 - \frac{M}{r} \right)^2 dt^2 - \left( 1 - \frac{M}{r} \right)^{-2} dr^2 - r^2 (d\theta ^2 + \sin ^2 \! \theta \; d\varphi ^2)
\end{equation}

Hereafter, we also call ``harmonic'' such a radial coordinate which, by means of the standard relations (\ref{1}), leads to the harmonic coordinates $X^1, X^2, X^3$. Equation (\ref{12}) implies the general harmonic $R$ to be
\begin{equation}\label{20}
R = r - M + K \frac{M^3}{(r-M)^2},
\end{equation}
where we put $C=1$ in (\ref{12}) so that by means of (\ref{1}) we get Minkowski coordinates at infinity. Equation (\ref{20}) is a cubic equation for $(r-M)$.  Its real root reads
\begin{eqnarray}\label{21}
r-M & = & \frac{1}{3} \left[ R + \frac{R^2}{S} + S \right],\\
S & \equiv & \left[ -\frac{27}{2} KM^3 + R^3 + \frac{3}{2}\sqrt{ 3KM^3(27 KM^3 - 4R^3) } \right]^{1/3}.\nonumber
\end{eqnarray}
Using this we can convert the metric (\ref{19}) explicitly into general harmonic coordinates. The resulting metric gets involved due to the complicated form of (\ref{21}); however, for our purposes it is sufficient to transform the metric (\ref{19}) into the following form:
\begin{eqnarray}\label{22}
ds^2 & = & \left( \frac{1-\mathcal{K}}{1+\frac{M}{R}-\mathcal{K}} \right) ^2 dt^2 - \left( \frac{1+\frac{M}{R}-\mathcal{K}}{1-3\mathcal{K}} \right)^2 dR^2 \nonumber \\
&& - \left( 1+\frac{M}{R} -\mathcal{K} \right)^2 R^2 (d\theta ^2 + \sin ^2 \! \theta \; d \varphi^2),
\end{eqnarray}
where the function
\begin{equation}\label{23}
\mathcal{K}(R) = \frac{KM^3}{R[r(R)-M]^2}
\end{equation}
is given explicitly only if we express $r(R)-M$ using (\ref{21}).

Now consider an interior solution (\ref{18}) in harmonic coordinates, in which $\{X, Y, Z\}$ are replaced by $\{R, \theta, \varphi\}$ using (\ref{1}); and $U = U(R)$ owing to the spherical symmetry:
\begin{equation}\label{24}
ds^2 = U^{-2} dt^2 - U^2 [dR^2 + R^2 (d\theta^2 + \sin ^2 \! \theta \; d\varphi^2)].
\end{equation}
We wish to match (\ref{24}) with (\ref{22}) at some radius $R_0$. Comparing the metric coefficients at $R=R_0$, we immediately see that the metric can be continuous across $R=R_0$ only if the constant $K = 0$. Therefore, in contrast to the vacuum Schwarzschild solution joined to the Schwarzschild interior solution as discussed in \cite{Li1}, \cite{MMR}, in the case of electrically counterpoised dust producing an extreme Reissner-Nordstr\"om solution outside, it is a \textit{necessary condition} for having ``admissible coordinates in the sense of Lichnerowicz'' that the exterior harmonic coordinate is the ``conventional one'' $R= r-M$, so that the second term in Eq. (\ref{15}) is absent. Various solutions are known explicitly that are described in global harmonic coordinates in which the metric is $C^1$ across the boundary. For example, to demonstrate the existence of globally regular solutions whose exteriors are near that of an extreme black hole and the redshift of radiation emitted at the centre can be arbitrarily large at infinity, Bonnor chose (with integer $n>1$) 
\begin{equation}\label{25}
U_I = 1 + \frac{M}{R_0} + \frac{M(R_0^n - R^n)}{nR_0^{n+1}}.
\end{equation}

The spheres of electrically counterpoised dust thus represent simple explicit examples of material bodies and their fields described in a completely regular way in the global harmonic coordinates. Still, a question arises whether another type of harmonic coordinates than those used in (\ref{24}) cannot be introduced \textit{inside} the spheres of electrically counterpoised dust. To tackle this equation we employ a simple interior solution which Bonnor gave in 1965 \cite{Bo1} with the square of $U_I(R)$ given by
\begin{equation}\label{26}
U^2_I = \frac{1}{A_0 + B_0R^2},
\end{equation}
where
\begin{equation}\label{27}
A_0 = [R_0/(R_0 + M)]^3,\;\;\; B_0 = M/(R_0 + M)^3
\end{equation}
are constants, $R$ is the usual isotropic coordinate entering the metric (\ref{24}).

With $U_I$ given by (\ref{26}), the metric (\ref{24}) can easily be transformed into Schwarz\-schild coordinates:
\begin{equation}\label{28}
ds^2 = \frac{A_0}{1-B_0r^2}dt^2 - \frac{1}{(1-B_0r^2)^2}dr^2 - r^2 (d\theta^2 + \sin^2 \! \theta \; d \varphi^2);
\end{equation}
the radial coordinates are connected by simple relations
\begin{equation}\label{29}
R = \sqrt{A_0} r/\sqrt{1-B_0r^2}, \;\;\;\;  r=R/\sqrt{A_0+B_0R^2}.
\end{equation}
It is easy to see that although the Schwarzschild-type coordinates are simple and appropriate for various purposes, they are not ``admissible in the sense of Lichnerowicz'' because the metrics (\ref{19}) and (\ref{28}) are not $C^1$ across the boundary $r=r_0$ corresponding to $R=R_0$. From the metric (\ref{28}) we can read off the explicit forms of the functions $a(r)$ and $\alpha(r)$ entering the differential equation (\ref{12}) for the harmonic coordinate $R$. Equation (\ref{12}) can be solved; the general solution is of the form

\begin{equation}\label{30}
R = C_1\rho + C_2 \frac{1}{\rho^2}, \;\;\; \rho\equiv\frac{\sqrt{B_0 }r}{\sqrt{1-B_0r^2}},
\end{equation}
$C_1, C_2$ are constants. Comparing this result, in which we put $C_2 = 0$ and $C_1 = \sqrt{A_0}/\sqrt{B_0}$ with $R$ given in Eq. (\ref{29}), this corresponds clearly to the standard harmonic coordinates; in these coordinates the interior metric is given by (\ref{24}) with $U_I$ determined by (\ref{26}). From Eq. (\ref{30}) we see, however, that there exists another possible choice for isotropic coordinates in the interior, namely $R = C_2 (1-B_0r^2)/B_0r^2$. Such $R$ is not a very ``natural'' radial coordinate since the origin $r=0$ is now mapped into $R = \infty$; nevertheless, it is the harmonic radial coordinate. Putting $C_1 = 0$, $C_2 = F/B_0$, metric (\ref{28}) becomes
\begin{eqnarray}\label{31}
ds^2 & = & \frac{A_0F}{\sqrt{B_0}R} \left( 1 + \sqrt{B_0}R / F \right) dt^2 - \left( \frac{1}{1 + \sqrt{B_0}R/F}\right) \frac{1}{4B_0R^2}dR^2 \nonumber \\
&& - \frac{1}{B_0(1+\sqrt{B_0}R/F)} (d\theta^2 + \sin^2 \! \theta \; d \varphi^2).
\end{eqnarray}
Can these harmonic coordinates become ``admissible'' across the boundary in the sense of Lichnerowicz? Amazingly, they can, provided that in the exterior region also an ``unconventional'' harmonic $R$, given by (\ref{15}) with $C = 0$, is used. Then
\begin{equation}\label{32}
R = KM^3/(r-M)^2,
\end{equation}
so that ``external'' physical infinity $(r\rightarrow \infty)$ is mapped into the origin $R = 0$. From (\ref{32}) one can easily express $r$ and find the exterior metric (\ref{19}) to become
\begin{eqnarray}\label{33}
ds^2 & = & \frac{KM^3/R}{(M+\sqrt{KM^3/R})^2} dt^2 - (M + \sqrt{KM^3/R})^2 \frac{1}{4R^2} dR^2 \nonumber \\
&& - (M + \sqrt{KM^3/R})^2 (d\theta ^2 + \sin ^2 \! \theta \; d\varphi^2).
\end{eqnarray}
Let us note that the verification which we made of the harmonic character of the coordinates in (\ref{31}) and (\ref{33}) is not immediate even when using the generalised (background covariant) forms (\ref{7}), (\ref{8}). However, it would be much more difficult if first the transformation (\ref{1}) to Cartesian-type coordinates is made and then non-covariant conditions (\ref{3}) are checked because in $\{ X, Y, Z\}$ coordinates the metrics (\ref{31}), (\ref{33}) become non-diagonal.

Now before matching the exterior and interior metrics at $R = R_0^*$ (we use the star  here to distinguish this value from $R_0$ in the standard harmonic coordinates) we have to match the coordinates (\ref{32}) and (\ref{30}) (with $C_1 = 0$, $C_2 = F/\sqrt{B_0})$ which implies 
\begin{equation}\label{34}
R^*_0 = \frac{KM^3}{(r_0 - M)^2} = \frac{F}{\sqrt{B_0}} \frac{(1-B_0r^2_0)}{B_0r^2_0}.
\end{equation}
Here $r_0$ is the Schwarzschild radius of the boundary, related to $R_0$ in the original harmonic coordinates just by $r_0 = R_0 + M$ so that the constants (\ref{27}) become $A_0 = (1-M/r_0)^3$ and $B_0 = M/r^3_0$. Then the relation (\ref{34}) determines the ratio of constants $K$ and $F$:
\begin{equation}\label{35}
\frac{K}{F} = \left( \frac{r_0}{M} - 1 \right)^3 (r_0/M)^{3/2}.
\end{equation}
Expressing $A_0$, $B_0$ in (\ref{31}) in terms of $r_0$, $M$ and comparing the metrics (\ref{31}), (\ref{32}) at $R^*_0$, we find out that both metrics and their first derivatives are continuous at $R = R^*_0$ provided that the relation (\ref{35}) holds. Hence, indeed, we found another set of harmonic coordinates which cover globally the sphere of electrically counterpoised dust and its exterior field and in which the metric is $C^1$ at the boundary. However, these harmonic coordinates become pathological at the origin and at infinity so that the first (standard) type of the harmonic coordinates in which the metric is given by Eq. (\ref{24}) everywhere are, of course, to be preferred.

\section{Stationary spacetimes with disc sources in global harmonic coordinates}

We now wish to give examples of stationary spacetimes with sources---though only with 2-dimensional matter distributions, and thus with no ``interior region''---de\-scribed in global harmonic coordinates. 

\subsection*{Conformastationary metrics}

The conformastationary Einsjtein-Maxwell fields discopiuypyvered by Neugebauer, Perj\'es, and Israel and Wilson around the beginning of 1970's (see, e.g., \cite{SKM} for references and more details) represent an interesting generalisation of the conformastatic metrics discussed in the preceding  section. As the first example, we consider conformastationary metrics with axial symmetry for which we constructed disc sources some time ago \cite{KBL1}. The discs are made of rotating charged dust but hoop tensions are necessary to balance the centrifugal forces induced by the motion. (This is at  variance with the statement in \cite{SKM} that the conformastationary solutions are the exterior fields of charged spinning sources in equilibrium under their mutual electrodynamic and gravitational forces.)

In the cylindrical-type coordinates the metric can be written as (see \cite{KBL1}, Eq. (2.2))
\begin{equation}\label{36}
ds^2 = f(dt - {\mathcal{A}}Rd\varphi)^2 - f^{-1} (dR^2 + dz^2 + R^2 d\varphi^2),
\end{equation}
where $f$ and $\mathcal{A}$ are functions of $(R,z)$. Putting $f = (VV^*)^{-1}$, where $\nabla^2 V = 0$, with $V$ complex, the electromagnetic field is given by $E + i H = \nabla (1/V)$ and ${\mathcal{A}}$ in (\ref{36}) is a solution of $\nabla \times {\mathcal{A}} = i(V\nabla V^* - V^* \nabla V)$. In \cite{KBL1}, in Eqs. (2.3)--(2.5), both the contravariant metric and the determinant $g$ are written down (there is a misprint in $g_{33}$ in (2.3)---the bracket should not be squared). We employ these expressions in the generalised harmonic conditions (\ref{7}), (\ref{8}). Regarding (\ref{36}), it is natural to take as the background metric in flat space $\overline{\mathcal{M}}$, the metric in cylindrical coordinates, i.e., Eq. (\ref{36}) with $f=1$, ${\mathcal{A}} = 0$. It is then easy to check that putting $\{ X^\mu = t, R, z, \varphi \}$, the generalised harmonic conditions (\ref{8}) are satisfied for $\mu = 0$ and $\mu = 3$, but, in fact, they are also satisfied for $\mu = 1$ and $\mu = 2$, as straightforward calculations show. Hence introducing coordinates $\{X, Y, Z\}$ by
$$X = R \cos \varphi, \;\;\;Y = R \sin \varphi, \;\;\;  Z = z,$$
the metric (\ref{36}) becomes
\begin{equation}\label{37}
ds^2 = f \left[ dt - {\mathcal{A}} R^{-1}(-YdX + XdY) \right]^2 - f^{-1} (dX^2 + dY^2 + dZ^2),
\end{equation}
where $\{t, X, Y, Z \}$ are harmonic, $R = \sqrt{X^2 + Y^2}$. Notice that no relation between ${\mathcal{A}}$ and $f$ was used to prove the harmonicity of these coordinates. The metric (\ref{37}) can describe various physical situations (for a simple example, see \cite{KBL1}, Section 8; for a conformastationary Kerr-Newman metric, see \cite{SKM}, Section 21.1.3).

As mentioned above, these axisymmetric conformastationary electrovacuum spacetimes can be produced by 2-dimensional discs made of rotating charged dust in steady motion and the tensions needed to guarantee equilibrium. These discs are of an infinite extent but they have finite total mass and charge which are equal (in geometrical units). It is of interest to ask what is the gyromagnetic ratio for these objects. It is defined as
\begin{equation}\label{38}
g = \frac{2MJ_e}{eJ_M},
\end{equation}
where $J_e$ is the total magnetic moment and $J_M$ the total angular momentum. At the end of \cite{KBL1} we found the gyromagnetic ratio for the discs  to be equal to 1 (as it is for ``classical'' objects). However, we defined the gyromagnetic ratio erroneously in \cite{KBL1} omitting the factor 2 in the definition (\ref{38}) (see below Eq. (8.23)). Indeed, the gyromagnetic ratio of sources producing conformastationary  metrics is equal to 2, like it is for charged, rotating (Kerr-Newman) black holes and as it is expected to be equal approximately for other highly relativistic objects (see \cite{PF}, \cite{NO} for recent studies of this problem).

\subsection*{The Kerr metric}
A transformation bringing the Kerr metric into harmonic coordinates was described in \cite{Ru}, but no explicit results for the metric were given there. (Curiously, the author of \cite{Ru} is one of the coauthors of a recent work \cite{AMR} in which the Kerr metric is transformed into harmonic coordinates only asymptotically near infinity, but Ref. \cite{Ru} is not referred to in \cite{AMR}.) Independently of \cite{Ru}, the authors of \cite{AN} transformed the Kerr metric into harmonic coordinates using a transformation by H.-G. Ding. They do give the metric explicitly (all 10 components are non-vanishing) but several of the $g_{\mu\nu}$'s represent expressions covering almost half a page in print. Nobody seemed to have checked whether this is indeed the Kerr metric in harmonic coordinates and the only application until now appears to be given in the same work \cite{AN}, where it was used to calculate the energy density for the Kerr spacetimes as determined from the Einstein pseudotensor. Nevertheless, in principle, the situation with the Kerr metric in harmonic coordinates is similar to that of (much simpler) conformastationary spacetimes in harmonic coordinates: one can construct 2-dimensional disc sources which have the Kerr metric outside \cite{BiL}, \cite{PiL}. The discs producing the general Kerr-Newman metrics can also be constructed---they are briefly described in \cite{LZ}. The Kerr-Newman metric in harmonic coordinates appears to be of interest in numerical relativity (see \cite{CO} for a recent review), but its explicit form (as given in [21]) does not seem to be noticed and no explicit form to be given by numerical relativists.

\section*{Acknowledgements}
We are grateful to the Institute of Astronomy, University of Cambridge and to Donald Lynden-Bell for their very kind hospitality. We thank Donald Lynden-Bell for inspiring remarks and Martin \v{Z}ofka for his help with the manuscript. J.B. acknowledges also discussions with Gerhard Sch\"{a}fer and support from the grant GA\v{C}R 202/02/0735 of the Czech Republic.

\end {document}